# Comment

**Stephen Senn**

I have always felt very guilty about Harold Jeffreys's *Theory of Probability* (referred to as *ToP*, hereafter). I take seriously George Barnard's injunction (Barnard, 1996) to have some familiarity with the four great systems of inference. I also consider it a duty and generally find it a pleasure to read the classics, but I find Jeffreys much harder going than Fisher, Neyman and Pearson fils or De Finetti. So I was intrigued to learn that Christian Robert and colleagues had produced an extensive chapter by chapter commentary on Jeffreys, honored to be invited to comment but apprehensive at the task.

Reading Robert et al.'s insightful commentary has sent me back to Jeffreys. Like them, what I am familiar with is the third edition (as corrected in 1966) and I have a rather battered copy with pages heavily annotated in pencil. My habit is to put a marginal vertical line against important passages that merit attention and a question mark where I don't understand. There are lots of both in my copy of Jeffreys.

The commentary by Roberts et al. is a tour de force. Only statisticians with complete familiarity with Bayesian methods and a deep understanding of its many forms could have produced it. It in no way detracts from my admiration for what the authors have achieved to have to admit that my opinion of Jeffreys is unchanged. *ToP* is full of brilliant insights and I return from it convinced that the man was a genius. However, I also think that to any outsider, the theory outlined *as a whole* will appear to be a bit of a mess.

As a small example of one of these insights, consider the discussion of "Artificial Randomization" in Section 4.9, not really covered by Robert et al. Among many interesting points, Jeffreys notes that if a $5 \times 5$ Latin Square in agriculture is analyzed using the methods proposed by Fisher, then the row and column totals have eight degrees of freedom assigned to them and hence that the polynomial equivalent is a quartic in the row and the column positions but with no cross-product terms, which would be a very strange function.

*Stephen Senn is Professor, Department of Statistics, University of Glasgow, Glasgow G12 8QW, Scotland e-mail: stephen@stats.gla.ac.uk.*

However, perhaps the most important insight in *ToP* concerns the necessity for a prejudice in favour of simpler theories if one wishes to try and rescue the Laplacian proposal of insufficient reason. I was once told by Peter Freeman that when he and Dennis Lindley interviewed Harold Jeffreys and asked him what he considered his greatest scientific achievement, they were stunned when he replied that it was the invention of the significance test. Thus Chapter V of *ToP* (reviewed by Roberts et al. in Section 6) is the one he regarded as being the most important.

There is a very interesting passage in a letter of Jeffreys to Fisher of 1 March 1934. [This correspondence forms pages 149–161 of Henry Bennett's edited correspondence of Fisher (Bennett, 1990) but is also available on the web in facsimile at the very useful site maintained by the University of Adelaide at http://digital.library.adelaide.edu.au/coll/special//fisher/.] The letter is part of a series initiated by the fact that papers of theirs that were due to appear in the *Proceedings of the Royal Society*. (Later in this correspondence, on 10 April, Jeffreys raises Newman's tramcar problem to which Robert et al. refer in Section 5.3.) John Aldrich (Aldrich, 2005), in an article I strongly recommend to any interested in Harold Jeffreys, has identified this period as being crucial to the statistical education of Jeffreys who was, it seems, long unaware that the biologists had something to teach the physicists.

Fisher and Jeffreys had been invited to take account of each other's submissions and were discussing what modification each should make (if any) to accommodate the other's position. The exchange is interesting because Jeffreys proves himself to be a fair match for Fisher and it is a tribute to the respect that Fisher clearly had for him that despite the fact that Jeffreys is occasionally rather cheeky to Fisher (suggesting, e.g., that if Fisher had chosen to justify likelihood in terms of work by Jeffreys and Dorothy Wrinch he would have been on strong ground), Fisher, who was sometimes irascible in correspondence, never loses his temper and even later proposes to moderate in the published commentary the terms in which he describes Jeffreys's theory.

The passage on page 3 of the letter of 1 March reads:

> I got as far as I could with the principle of non-sufficient reason, but it turns out in some cases to give answers quite contrary to general belief. E.G. as Broad pointed out, it will never give a reasonable probability to a general law of the form 'all crows are black.'

The reference is to the Cambridge philosopher, C. D. Broad (1887–1971) and to the fact that he had noted (Broad, 1918) in *Mind* that if Laplace's Law of Succession (discussed by Robert et al. in Section 4.2) applies and if $m$ counters are drawn at random from a bag known to contain $n$ counters and all drawn are observed to be white, then the probability that all $n$ are white is $(m+1)/(n+1)$ and hence that even if several logical difficulties in applying the "law" are ignored, it will never provide a means of proving the probable truth of any scientific law, for which $n$ must be effectively infinite. [See Aldrich, 2005 for a discussion, and also Chapter 4 of *Dicing with Death* (Senn, 2003) for a heuristic explanation.]

Subsequently, on page 4 of the original letter, Jeffreys writes:

> The principle of non-sufficient reason is intended simply to serve as an expression of lack of prejudice.... But in these cases of general laws there seems to be prejudice; I cannot help it, but there is a general belief in the possibility of establishing quantitative laws by experience, and I am not prepared to say that the general belief is wrong.

An interesting feature of this is that Jeffreys effectively admits that the necessity of being able to assert the conclusion is the justification of the premise. The solution that he found, as Robert et al. discuss, is to give a lump of probability to the precise form of the law. I think that this was a touch of genius, necessary to rescue the Laplacian formulation. Paradoxically, I suspect, however, that this particular aspect of Jeffreys's program is much less used than the other main feature, namely the use of improper prior distributions, and it is odd that he did not draw the conclusion that the principle of non-sufficient reason is generally unusable. It seems to me that whatever philosophical difficulties Jeffreys may have had in accepting Fisher's frequency limit view of probability ought to apply in spades to such improper "probabilities."

I have two very minor criticisms of Robert et al. The first is that one should be careful when talking about *Bernoulli*. I have a personal probability close to 1 that the Bernoulli of Section 2.3 is Daniel (1700–1782) and a similar probability that the Bernoulli of Section 3 is the same as that in the footnote to Section 9 and hence is his uncle James (1654–1705). It would have been helpful to the reader to have them distinguished. The second is that if one refers to Lindley's paradox (Lindley, 1957), one should also refer to Bartlett's correction (Bartlett, 1957), not least because when the correction is applied the paradox is seen to be not quite so automatic: two Bayesians could strongly disagree with each other. Thus, to adapt the rhetoric of Jeffreys's criticism of significance tests on page 385 of *ToP*, if one takes the paradox as being a reason for rejection Fisherian significance tests:

> it would require that a procedure is dismissed because, when combined with information which it doesn't require and which may not exist, it disagrees with a procedure that disagrees with itself (Senn, 2001).

Finally, let me say that although there are some aspects of *ToP* which do not enthuse me, it is nonetheless full of startling and brilliant insights and all the more welcome because at the time it appeared it provided a fresh and distinct alternative to a developing orthodoxy. Statisticians like me have every reason to be grateful to Robert et al. for helping us to obey Barnard's injunction.

## ACKNOWLEDGMENT

This work was undertaken while a guest of the CREST laboratory of ENSAI in Rennes, France and I am very grateful to my hosts for the invitation.

COMMENT 3